\documentclass[11pt]{article}

\newsavebox{\foobox}
\newcommand{\slantbox}[2][0]{\mbox{%
        \sbox{\foobox}{#2}%
        \hskip\wd\foobox
        \pdfsave
        \pdfsetmatrix{1 0 #1 1}%
        \llap{\usebox{\foobox}}%
        \pdfrestore
}}
\newcommand\unslant[2][-.25]{\slantbox[#1]{$#2$}}

\newcommand{\mmu}{\text{\unslant\mu}}
\newcommand{\mpi}{\text{\unslant[-.18]\pi}}
\newcommand{\mdelta}{\text{\unslant[-.18]\delta}}

\usepackage[left=2cm, right=2cm, top=2.5cm, bottom=2.5cm]{geometry}
\usepackage[x11names]{xcolor}
\usepackage{fancyhdr, amssymb, cancel, amsmath, graphicx, pgfplots, tikz}
\usepackage{isomath}
\usepackage{cite}

\usetikzlibrary{shadows}

\newcommand{\stylecolor}{blue!50!black}

\usepackage[labelfont={bf,sf, color=\stylecolor}, margin={1.5cm,0cm}]{caption}

\usepackage[colorlinks=true, urlcolor=\stylecolor!70!white, linkcolor=\stylecolor, citecolor=\stylecolor!70!white, hyperindex=true, linktocpage=true]{hyperref}

\usepackage[explicit]{titlesec}

\newcommand*\sectionlabel{}
\titleformat{\section}
  {\gdef\sectionlabel{}
   \Large\bfseries\scshape}
  {\gdef\sectionlabel{\thesection }}{0pt}
  {\begin{tikzpicture}[remember picture,overlay]
       \end{tikzpicture}
  }
\titlespacing*{\section}{0pt}{0pt}{0pt}

\newcommand*\subsectionlabel{}
\titleformat{\subsection}
  {\gdef\subsectionlabel{}
   \large\bfseries\scshape}
  {\gdef\subsectionlabel{\thesubsection  }}{0pt}
  {\begin{tikzpicture}[remember picture]
    \draw (-0.15, 0) node[left] {\color{\stylecolor} \textsf{\subsectionlabel}};
    \draw (0.15, 0) node[right] {\color{\stylecolor} \textsf{#1}};
    \fill[color=\stylecolor] (-0.05, -0.23) rectangle (0.05, 0.23);
       \end{tikzpicture}
  }
\titlespacing*{\subsection}{-4pt}{10pt}{0pt}

\newcommand*\subsubsectionlabel{}
\titleformat{\subsubsection}
  {\gdef\subsubsectionlabel{}
   \bfseries\scshape}
  {\gdef\subsubsectionlabel{\thesubsubsection.\ \  }}{0pt}
  {\begin{tikzpicture}[remember picture]
    \draw (0, 0) node[left] {\color{\stylecolor} \textsf{\subsubsectionlabel}};
    \draw (0, 0) node[right] {\color{\stylecolor} \textsf{#1}};
       \end{tikzpicture}
  }
\titlespacing*{\subsubsection}{-4pt}{7pt}{0pt}

\pgfplotsset{every axis legend/.append style={at={(1.02,1)},anchor=north west}}

\graphicspath{{figures/}}

\begin{document}

\allowdisplaybreaks

\pagestyle{fancy}
\renewcommand{\headrulewidth}{0pt}
\fancyhead{}

\fancyfoot{}
\fancyfoot[C] {\textsf{\textbf{\thepage}}}

\begin{equation*}
\begin{tikzpicture}
\draw (\textwidth, 0) node[text width = \textwidth, right] {\color{white} easter egg};
\end{tikzpicture}
\end{equation*}

\begin{equation*}
\begin{tikzpicture}
\draw (0.5\textwidth, -3) node[text width = \textwidth] {\huge \textsf{\textbf{Dyakonov-Shur instability across the ballistic- \\ \vspace{0.07in} to-hydrodynamic crossover}} };
\end{tikzpicture}
\end{equation*}
\begin{equation*}
\begin{tikzpicture}
\draw (0.5\textwidth, 0.1) node[text width=\textwidth] {\large \color{black} \textsf{Christian B. Mendl}$^{\color{\stylecolor} \mathsf{a,b,c}}$ \textsf{and Andrew Lucas}$^{\color{\stylecolor} \mathsf{c}}$};
\draw (0.5\textwidth, -1.5) node[text width=\textwidth] {$^{\color{\stylecolor} \mathsf{c}}$ \small\textsf{Department of Physics, Stanford University, Stanford, CA 94305, USA}};
\draw (0.5\textwidth, -1) node[text width=\textwidth] {$^{\color{\stylecolor} \mathsf{b}}$ \small\textsf{Stanford Institute for Materials and Energy Sciences, SLAC National Laboratory, Menlo Park, CA 94025, USA}};
\draw (0.5\textwidth, -0.5) node[text width=\textwidth] {$^{\color{\stylecolor} \mathsf{a}}$ \small\textsf{Technische Universit\"at Dresden, Institute of Scientific Computing, Zellescher Weg 12-14, 01069 Dresden, Germany}};
\end{tikzpicture}
\end{equation*}
\begin{equation*}
\begin{tikzpicture}
\draw (0, -13.1) node[right, text width=0.5\paperwidth] {\texttt{christian.mendl@​tu-dresden.de, ajlucas@stanford.edu}};
\draw (\textwidth, -13.1) node[left] {\textsf{\today}};
\end{tikzpicture}
\end{equation*}
\begin{equation*}
\begin{tikzpicture}
\draw[very thick, color=\stylecolor] (0.0\textwidth, -5.75) -- (0.99\textwidth, -5.75);
\draw (0.12\textwidth, -6.25) node[left] {\color{\stylecolor} \textsf{\textbf{Abstract:}}};
\draw (0.53\textwidth, -6) node[below, text width=0.8\textwidth, text justified] {\small We numerically solve semiclassical kinetic equations and compute the growth rate of the Dyakonov-Shur instability of a two-dimensional Fermi liquid in a finite length cavity. When electron-electron scattering is fast, we observe the well-understood hydrodynamic instability, and its disappearance due to viscous dissipation. When electron-electron scattering is negligible, we find that the instability re-emerges for certain boundary conditions, but not for others. We discuss the implications of these findings for experiments.};
\end{tikzpicture}
\end{equation*}

\tableofcontents

\begin{equation*}
\begin{tikzpicture}
\draw[very thick, color=\stylecolor] (0.0\textwidth, -5.75) -- (0.99\textwidth, -5.75);
\end{tikzpicture}
\end{equation*}

\titleformat{\section}
  {\gdef\sectionlabel{}
   \Large\bfseries\scshape}
  {\gdef\sectionlabel{\thesection }}{0pt}
  {\begin{tikzpicture}[remember picture]
    \draw (0.2, 0) node[right] {\color{\stylecolor} \textsf{#1}};
    \draw (0.0, 0) node[left, fill=\stylecolor,minimum height=0.27in, minimum width=0.27in] {\color{white} \textsf{\sectionlabel}};
       \end{tikzpicture}
  }
\titlespacing*{\section}{0pt}{20pt}{5pt}

\section{Introduction}

The spontaneous generation of terahertz radiation is an important yet challenging problem in applied physics \cite{kitaeva}. An interesting proposal is to generate terahertz radiation through the Dyakonov-Shur (DS) instability of a two-dimensional electron gas (2DEG) \cite{DS, DS2, DS05}. This instability occurs in a uniform flow of current through the 2DEG, subject to non-standard, but experimentally achievable, boundary conditions. In the $xy$-plane, we consider an infinite strip of 2DEG of width $L$ ($0 \le x \le L$). A uniform, small, background current density $J_x > 0$ is pushed through the strip, and we fix density fluctuations to vanish at $x=0$, and current fluctuations to vanish at $x = L$. Assuming homogeneity in the $y$-direction, one finds that for small currents $J_x$, an instability arises. Spontaneous fluctuations in density and current of amplitude $\epsilon$ at time $t=0$ grow to amplitude $\epsilon \mathrm{e}^{\gamma t}$ at time $t$. In terms of the fluid velocity $u_0 = J_x/\rho$, with $\rho$ the charge density of the 2DEG, \begin{equation}
\gamma \approx \frac{u_0}{L}.
\label{eq:gamma0}
\end{equation}

Some signatures of the DS instability have been found in experiment \cite{tauk, giliberti}, but a clear observation of the DS instability remains challenging. Perhaps one reason is that the original proposal \cite{DS} for the instability was in a \emph{hydrodynamic regime} \cite{lucasreview2}, where electrons collide with other electrons at a rate $1/\tau_{\mathrm{ee}}$ much larger than the rate $1/\tau_{\mathrm{imp}}$ of electron-impurity/phonon or umklapp collisions. With a few exceptions \cite{molenkamp, mackenzie, bandurin, crossno, levitov1703}, most electron liquids have \emph{not} been experimentally observed in a hydrodynamic regime. However, an interesting assertion is that the DS instability also exists in a ballistic limit where $\tau_{\mathrm{ee}} \rightarrow \infty$ and $\tau_{\mathrm{imp}} \rightarrow \infty$ \cite{DSkin}. If the hydrodynamic limit is not necessary, then the DS instability should be observable in a much larger set of 2DEGs and temperature ranges. The difficulty of observing the DS instability would be even more puzzling.

A quick check of this assertion is to compute the viscous correction to $\gamma$ \cite{DS, DS2, rudin}: \begin{equation}
\gamma = \frac{u_0}{L} - \frac{\mpi^2 \nu}{8L^2}, \label{eq:gamma1}
\end{equation}
with $\nu \sim v_{\mathrm{F}}^2\tau_{\mathrm{ee}}$ the dynamical viscosity, and $v_{\mathrm{F}}$ the Fermi velocity. For simplicity in \eqref{eq:gamma1}, and throughout this letter, we take $\tau_{\mathrm{imp}} \rightarrow \infty$. The hydrodynamic limit corresponds to $v_{\mathrm{F}}\tau_{\mathrm{ee}} \lesssim L$. If this inequality is saturated, we estimate that $\gamma \sim -(v_{\mathrm{F}}-u_0)/v_{\mathrm{F}}\tau_{\mathrm{ee}}$, which is expected to be negative. This simple calculation suggests that the DS instability could vanish if electron-electron interactions are weak enough.

In this letter we explicitly check the fate of the DS instability and numerically calculate $\gamma$ for a two-dimensional Fermi liquid, using a toy model of (quantum) kinetic theory, with suitable boundary conditions. When $v_{\mathrm{F}}\tau_{\mathrm{ee}}\ll L$, we observe quantitative agreement with \eqref{eq:gamma1}. When $v_{\mathrm{F}}\tau_{\mathrm{ee}}\gg L$, we find that the instability becomes somewhat sensitive to boundary conditions. For ``clean" boundaries with specular scattering, we numerically find that \begin{equation}
\gamma \approx \frac{u_0}{L} - \frac{\mpi^2 \nu }{8L^2} \dfrac{1}{1+\left(\dfrac{\mpi v_{\mathrm{s}}}{2L} \tau_{\mathrm{ee}}\right)^2} \label{eq:gammaapprox}
\end{equation}
approximates the instability growth rate. Here $v_{\mathrm{s}}$ is the speed of sound in the electron fluid. Hence, as $\tau_{\mathrm{ee}} \rightarrow \infty$, we recover \eqref{eq:gamma0}, in agreement with \cite{DSkin}. However, for ``dirty" boundary conditions with non-specular scattering, we numerically observe that $\gamma<0$ becomes possible as $\tau_{\mathrm{ee}} \rightarrow \infty$.   Our results demonstrate how boundary conditions on non-hydrodynamic modes could play an important role in suppressing the DS instability in experimental systems.



\section{Kinetic Theory}

We now turn to more quantitative details of our study. We compute the low temperature dynamics of an isotropic Fermi liquid in $d=2$ spatial dimensions, employing the model of \cite{molenkamp, levitov1607, lucas1612, levitov1612}. A thorough introduction to this model is given in Appendix \ref{app:kinetic}; here we summarize the key points. At low temperatures compared to the Fermi temperature, the most important semiclassical dynamics of a Fermi liquid correspond to the ``sloshing" of the Fermi surface itself. If we are only interested in dynamics on length scales large compared to the Fermi wavelength $\lambda_{\mathrm{F}}$, then it suffices to solve for the fermion distribution function $f(\mathbf{x},\mathbf{p})$. Heuristically, $f$ is the ``number density of quasiparticles of momentum $\mathbf{p}$ at position $\mathbf{x}$", and the Pauli exclusion principle forces $0\le f \le 1$ for electronic quasiparticles. It is useful to write $f$ as
\begin{equation}
f(\mathbf{x},\mathbf{p}) \approx n_{\mathrm{F}}(\epsilon(\mathbf{p}) - \mu - \Phi(\mathbf{x},\mathbf{p}))
\label{eq:f}
\end{equation}
with $n_{\mathrm{F}}(x) \approx \mathrm{\Theta}(-x)$ at low temperature and $\mathrm{\Theta}$ the Heaviside step function. If the perturbation $\Phi$ is small, 
\begin{equation}
f(\mathbf{x},\mathbf{p}) \approx f_{\mathrm{eq}}(\mathbf{p}) + \mdelta(\epsilon(\mathbf{p}) - \mu) \Phi(\mathbf{x},\mathbf{p}).
\end{equation}
If the Fermi surface is isotropic, and (for now) there is no background velocity ($u_0=0$), then the $\mdelta$ function above simply fixes $|\mathbf{p}| = p_{\mathrm{F}}$, and $\Phi$ may be parametrized by the angular component $\theta$ of $\mathbf{p}$:
\begin{equation}
\Phi = \Phi(\mathbf{x},\theta) = \sum_{n\in\mathbb{Z}} a_n(\mathbf{x},t) \mathrm{e}^{\mathrm{i}n\theta}. \label{eq:Phi}
\end{equation}
The harmonic $a_0$ is proportional to fluctuations in the number density of electrons, while $a_{\pm 1}$ correspond to the local density of $(x\pm \mathrm{i}y)$-momentum. In the toy model of \cite{molenkamp, levitov1607, lucas1612, levitov1612}, the dynamical time evolution of $\Phi$ is described by a Boltzmann equation in a relaxation time approximation \cite{bgk}: \begin{equation}
\partial_t \Phi + v_{\mathrm{F}}\cos(\theta) \partial_x \Phi = -\frac{1}{\tau_{\mathrm{ee}}}\mathsf{P}[\Phi],
\label{eq:boltzmann}
\end{equation}
where \begin{equation}
\mathsf{P}[\Phi] = \sum_{|n|\ge 2} a_n \mathrm{e}^{\mathrm{i}n\theta}.
\label{eq:PPhi}
\end{equation}
Due to our setup, we have assumed $\partial_y=0$. The term on the right hand side of \eqref{eq:boltzmann} is the linearized collision integral: it relaxes all harmonics of $\Phi$ that do not encode a conserved quantity. This model is not microscopically accurate \cite{ledwith1, ledwith2}, but correctly reproduces both $\tau_{\mathrm{ee}}=0$ and $\tau_{\mathrm{ee}}=\infty$ limits.

Our model does not account for electron-impurity scattering. Heuristically, if $\tau_{\mathrm{imp}}$ is the electron-impurity scattering rate, then $\gamma \rightarrow \gamma - 1/2\tau_{\mathrm{imp}}$ \cite{DSdis, crowne2}. High quality 2DEGs can reach $v_{\mathrm{F}}\tau_{\mathrm{imp}} \gtrsim 15 \; \mmu$m \cite{mackenzie, wang13}, which is larger than the typical device size.

For mathematical simplicity, we now take 
\begin{equation}
\epsilon(\mathbf{p}) = \frac{\mathbf{p}^2}{2m}.
\label{eq:gal}
\end{equation}
To account for background flow, we simply use Galilean invariance: $\partial_t \rightarrow \partial_t + u_0 \partial_x$ in \eqref{eq:boltzmann}.

In many experimentally realized 2DEGs, the Coulomb interactions are screened by conductors (``gates") a few nm above the sample. This causes an external force \cite{DS} 
\begin{equation}
\mathbf{F} = \frac{e^2}{C} \nabla n
\label{eq:Fgate}
\end{equation} 
on the electron liquid, analogous to a non-vanishing Landau parameter $\mathcal{F}_0$ \cite{lucasdassarma}. Here $C$ is the capacitance of the gates per unit area, and $n$ is the number density of electrons (note $n\propto a_0$). Looking for normal modes where $\Phi \sim \mathrm{e}^{-\mathrm{i}\omega t}$, \eqref{eq:boltzmann} generalizes to 
\begin{equation}
\mathrm{i}\omega\Phi = \big(u_0+ v_{\mathrm{F}}\cos(\theta)\big) \partial_x \Phi + \frac{2v_{\mathrm{g}}^2}{v_{\mathrm{F}}} \cos(\theta) \partial_x a_0 + \frac{1}{\tau_{\mathrm{ee}}}\mathsf{P}[\Phi],
\label{eq:boltzmannfin}
\end{equation}
with \begin{equation}
v_{\mathrm{g}}^2 = \frac{e^2n_0}{mC},
\label{eq:vgate}
\end{equation}
with $n_0$ the background electron density. $\gamma$ is given by $\max(\mathrm{Im}(\omega_*))$, where $\omega_*$ are the eigenvalues of \eqref{eq:boltzmannfin}, subject to suitable boundary conditions.

In the hydrodynamic limit, the DS instability is caused by sound waves with dispersion relation \begin{equation}
\omega \approx (u_0\pm v_{\mathrm{s}})k - \mathrm{i}\frac{\nu}{2}k^2,
\label{eq:disp}
\end{equation} 
with \begin{equation}
v_{\mathrm{s}} = \sqrt{\frac{v_{\mathrm{F}}^2}{2} + v_{\mathrm{g}}^2 }, \qquad \nu = \frac{v_{\mathrm{F}}^2\tau_{\mathrm{ee}}}{4}.
\label{eq:nudef}
\end{equation}
Neglecting the effects of gating leads to a universal speed of sound $v_{\mathrm{F}}/\sqrt{2}$ \cite{levitov1607, lucas1612}. In the limit where the dominant forces on electrons arise from the gate, $v_{\mathrm{g}} \gg v_{\mathrm{F}}$ and we recover the speed of sound of \cite{DS, DSkin}. Assuming $u_0 \ll v_{\mathrm{s}}$, we can estimate the growth rate $\gamma$ of the DS instability as follows. The DS boundary conditions amplify sound waves that scatter off of the fixed-current boundary. The rate of these scattering events $\sim v_{\mathrm{s}}/L$, and the amplification factor is $\sim u_0/v_{\mathrm{s}}$. A sound wave of any amplitude decays at a fixed rate, given in \eqref{eq:disp}, with $k \approx \mpi / 2L$. Adding the amplification rate and the viscous decay rate leads to \eqref{eq:gamma1}.

In the ballistic limit, a crude approximation is that the most important corrections to hydrodynamics can be accounted for by a frequency-dependent viscosity \cite{zaanen}: \begin{equation}
\nu(\omega) = \frac{v_{\mathrm{F}}^2\tau_{\mathrm{ee}}}{4(1-\mathrm{i}\omega \tau_{\mathrm{ee}})}.
\label{eq:nuomega}
\end{equation}
This equation appears qualitatively consistent with more microscopic calculations in graphene \cite{polini1506}, and can be derived by crudely truncating \eqref{eq:boltzmannfin} to a few harmonics (see Appendix \ref{app:kinetic}). Estimating that we must replace $\nu$ in \eqref{eq:disp} with $\mathrm{Re}(\nu(\omega))$, and approximating $\omega \approx \mpi v_{\mathrm{s}}/2L$ when evaluating $\nu(\omega)$, we obtain our \emph{heuristic} result \eqref{eq:gammaapprox}.

When $u_0=0$, we can also study the minimal quality factor $Q = \min_k [-\mathrm{Re}(\omega(k))/\mathrm{Im}(\omega(k))]$ of the waves.   Using the approximations of the previous paragraph, we estimate \begin{equation}
Q\approx \frac{4\tau_{\mathrm{ee}}v_{\mathrm{s}}^2}{\nu} \approx 8 + 16\frac{v_{\mathrm{g}}^2}{v_{\mathrm{F}}^2}.  \label{eq:Qmin}
\end{equation}
This is in qualitative agreement with the Q-factor reported recently in \cite{svintsov} in a similar model.

\section{Numerical Results}

For finite $\tau_{\mathrm{ee}}$, we calculate $\omega_*$ and $\gamma$ numerically by truncating \eqref{eq:Phi} to modes with $|n| \le n_{\mathrm{max}}$. Details of the numerical methods can be found in Appendix \ref{app:numerics}. 
The DS boundary conditions are \begin{subequations}\label{eq:DSBC}\begin{align}
0 &= a_0(0), \\
0 &= u_0 a_0(L) + \frac{v_{\mathrm{F}}}{2}(a_1(L)+a_{-1}(L)).
\end{align}\end{subequations}
Choosing the remaining boundary conditions on $a_n$ for $|n|\ge 2$ requires some more care. For example, the number of boundary conditions required by the truncated \eqref{eq:boltzmannfin} is $2n_{\mathrm{max}}$ when $u_0=0$, and $2n_{\mathrm{max}}+1$ otherwise. The final boundary condition at $u_0>0$ must be chosen so the $u_0 \rightarrow 0$ limit is not singular.\footnote{This issue is discussed in hydrodynamic language in \cite{crowne}; our resolution appears to be new.} We have found that the proper choice of this boundary condition is $a_2(0)+a_{-2}(0)=0$. A natural choice to fix the remaining $2n_{\mathrm{max}}-2$ boundary conditions, is to demand that, up to the three prior boundary conditions, $\Phi(\theta) = \Phi(\mpi - \theta)$, or $a_n = (-1)^n a_{-n}$. Physically, this boundary condition states that the contacts to the 2DEG are atomically ``clean": quasiparticles specularly reflect off of the boundary. Alternative ``dirty" boundary conditions are that incoming particles reflect back at a random (outgoing) angle. More details on the choice of boundary conditions is provided in Appendix \ref{app:bc}.

\begin{figure}[t]
\centering
\includegraphics[width=0.55\textwidth]{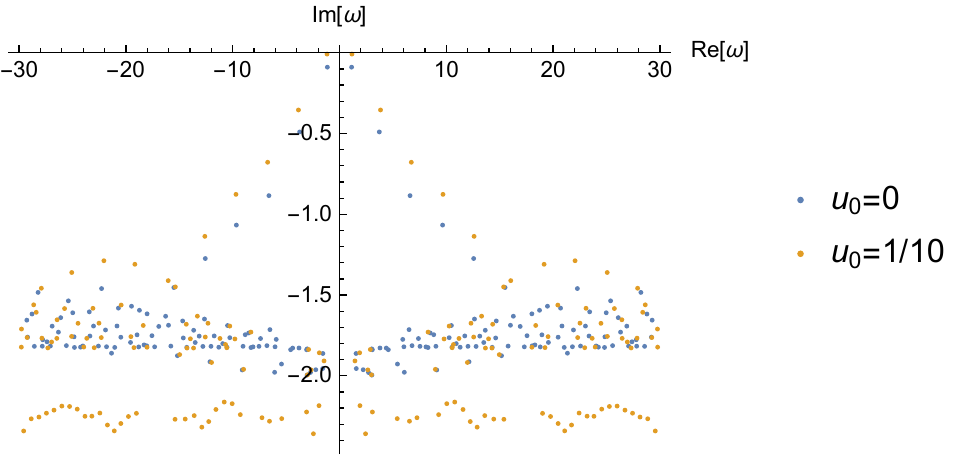}
\caption{The even part of the eigenvalue spectrum of \eqref{eq:boltzmannfin} with $v_{\mathrm{g}}=0$, $\tau_{\mathrm{ee}} = 1/2$ and DS boundary conditions, for two values of $u_0$. For small $\tau_{\mathrm{ee}}$, the instability arises exclusively in the hydrodynamic sound channel (points on the fictitious curve approaching $\omega = 0$). An infinite number of ballistic modes appears for $\mathrm{Im}(\omega) \approx -1/\tau_{\mathrm{ee}}$.}
\label{fig:eigs}
\end{figure}

For now, let us take clean boundary conditions, up to the caveats of the previous paragraph. We present the entire eigenvalue spectrum in Figure \ref{fig:eigs}, corresponding to fluctuations which are even under $y\rightarrow -y$ (the odd sector decouples). As expected, we observe that the DS instability is carried entirely by sound modes in the hydrodynamic limit, within the full kinetic theory. All non-hydrodynamic degrees of freedom have a finite decay rate: $\mathrm{Im}(\omega_*) \approx -\tau_{\mathrm{ee}}^{-1}$. In the infinite volume limit with $u_0=0$, \cite{lucasdassarma} has shown analytically that $\mathrm{Im}(\omega_*) \approx -\tau_{\mathrm{ee}}^{-1}$ for all non-hydrodynamic modes. In all plots in this letter, we work in units where $v_{\mathrm{F}}=L=1$; thus $\tau_{\mathrm{ee}} <1$ is ``hydrodynamic" and $\tau_{\mathrm{ee}} > 1$ is ``ballistic".

\begin{figure}[t]
\centering
\includegraphics[width=0.5\textwidth]{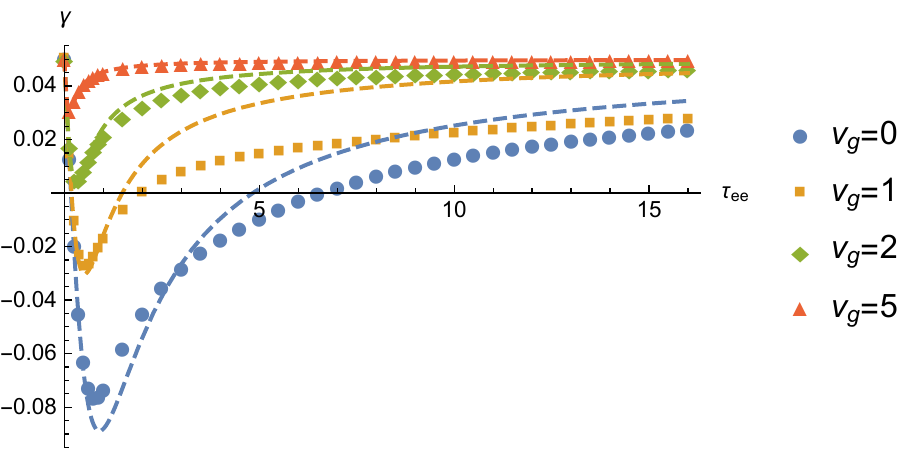}
\caption{$\gamma$ as a function of $\tau_{\mathrm{ee}}$, for $u_0 = 1/20$, various gate voltages $v_g$ and clean boundary conditions. An increasing gate voltage favors the instability. Solid markers show numerical data points, while the dashed line is our heuristic analytic result \eqref{eq:gammaapprox}.}
\label{fig:gammafunc}
\end{figure}

\begin{figure}[t]
\centering
\includegraphics[width=0.45\textwidth]{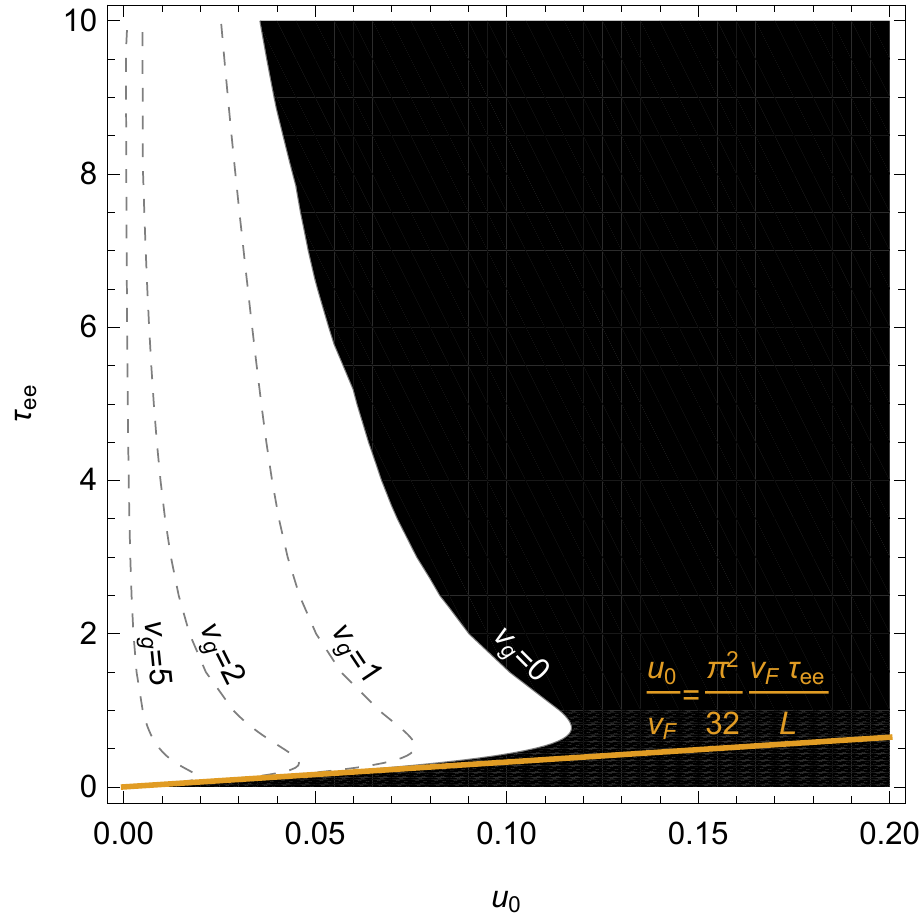}
\caption{Values of $u_0$ and $\tau_{\mathrm{ee}}$ where $\gamma>0$ ($\gamma<0$) are shown in black (white). Dashed lines show the regime of instability at finite $v_{\mathrm{g}}$. The gold line shows \eqref{eq:gamma1}.}
\label{fig:clean}
\end{figure}

Continuing to assume clean boundary conditions, we next compute $\gamma$ as a function of both $\tau_{\mathrm{ee}}$ and $v_{\mathrm{g}}$, for fixed $u_0>0$; the result is shown in Figure \ref{fig:gammafunc}. Regardless of $v_{\mathrm{g}}$, we find \eqref{eq:gamma1} universally in the hydrodynamic limit. Once $v_{\mathrm{F}}\tau_{\mathrm{ee}} \sim L$, we observe that $\gamma$ reaches a \emph{minimal value} $\gamma_{\mathrm{min}}$. For larger $\tau_{\mathrm{ee}}$, $\gamma$ increases as $\tau_{\mathrm{ee}}$ increases. In fact, we observe that for any $v_{\mathrm{g}}$, once $\tau_{\mathrm{ee}}$ is large enough, $\gamma>0$ (for these boundary conditions). The DS instability occurs in both the hydrodynamic and the ballistic limits, while possibly disappearing at the crossover between them, depending on $u_0$ and $v_{\mathrm{g}}$. Figure \ref{fig:gammafunc} also confirms that our heuristic estimate \eqref{eq:gammaapprox} captures the qualitative physics of the entire ballistic-to-hydrodynamic crossover. Figure \ref{fig:clean} gives an alternate perspective, showing where $\gamma$ is positive or negative as a function of $\tau_{\mathrm{ee}}$ and $u_0$. The ``lobe" shape where the instability disappears in Figure \ref{fig:clean} is equivalent to the dip in $\gamma(\tau_{\mathrm{ee}})$ observed in Figure \ref{fig:gammafunc}: the DS instability is most suppressed when $v_{\mathrm{F}}\tau_{\mathrm{ee}} \sim L$. Although one cannot  directly compare the minimal Q-factor in Figure \ref{fig:gammafunc} with \eqref{eq:Qmin}, as $u_0 >0$, we do observe that the width and magnitude of the dip  in  $\gamma$ both decrease as $v_{\mathrm{g}}$ increases, in agreement with (\ref{eq:Qmin}). Numerical data in Figure \ref{fig:gammafunc} is qualitatively consistent with a Q-factor $\gtrsim 10$, again in agreement with \eqref{eq:Qmin} and \cite{svintsov}.

We have numerically observed that $\gamma$ is insensitive to boundary conditions in the hydrodynamic limit. The ballistic limit, however, is sensitive to boundary conditions, and an accurate numerical computation of $\gamma$ can become quite challenging. In the collisionless limit $\tau_{\mathrm{ee}} \to \infty$, the equations \eqref{eq:boltzmannfin} for $\Phi(x,\theta)$ decouple at every $\theta$, and a uniform discretization $\theta_j = 2 \pi j/n_{\max}$ for $j = 0, 1, \dots, n_{\max}-1$ becomes more natural than a (spectral) harmonic truncation: see Appendix \ref{app:numerics_angular_discretization}. Thus the functions $\Phi(x,\theta_j)$ are only coupled via the boundary conditions. The DS conditions of vanishing density and current fluctuations at $x = 0$ and $x = L$ translate to $\sum_{j} \Phi(0,\theta_j) = 0$ and $\sum_{j} (u_0 + v_{\mathrm{F}}\cos(\theta_j)) \Phi(L,\theta_j) = 0$, respectively. Besides the DS conditions, we additionally use either clean boundary conditions $\Phi(x, \theta_j) = \Phi(x, \pi - \theta_j)$ at both ends, or a ``no-slip'' reflection $\Phi(0, \theta_j) = \Phi(0, \theta_j + \pi)$ on the left together with ``dirty'' boundary conditions at $x = L$, such that the distribution $\Phi(L, \theta_j)$ for ``outgoing'' angles $\theta_j$ (i.e., $u_0 + v_{\mathrm{F}}\cos(\theta_j) < 0$) is uniform. Intuitively, these dirty boundary conditions correspond to an atomically rough contact surface, upon which an incoming quasiparticle is equally likely to be scattered off of the boundary at any scattering angle.  A detailed description of the dirty boundary conditions is provided in Appendix~\ref{app:dirty_bc_theta}. Figure \ref{fig:ballistic} shows the numerically computed eigenvalue spectrum of the collisionless kinetic equation for these two variants of boundary conditions, at fixed $u_0$. We observe that for clean boundary conditions, the instability is present, while for no-slip -- dirty boundary conditions the instability is absent.

Our finding that dirty boundary conditions destroy the DS instability is consistent with \cite{satou}, which found that boundary conditions can effectively damp excitations in a finite length cavity. However, we have also demonstrated the existence of boundary conditions where the DS instability is recovered in the ballistic limit. For certain values of $u_0$ and $v_{\mathrm{g}}$, it is possible for the DS instability to persist for arbitrary electron-electron scattering times $\tau_{\mathrm{ee}}$, as depicted in Figure~\ref{fig:gammafunc}.

\begin{figure}[t]
\centering
\includegraphics[width=0.6\textwidth]{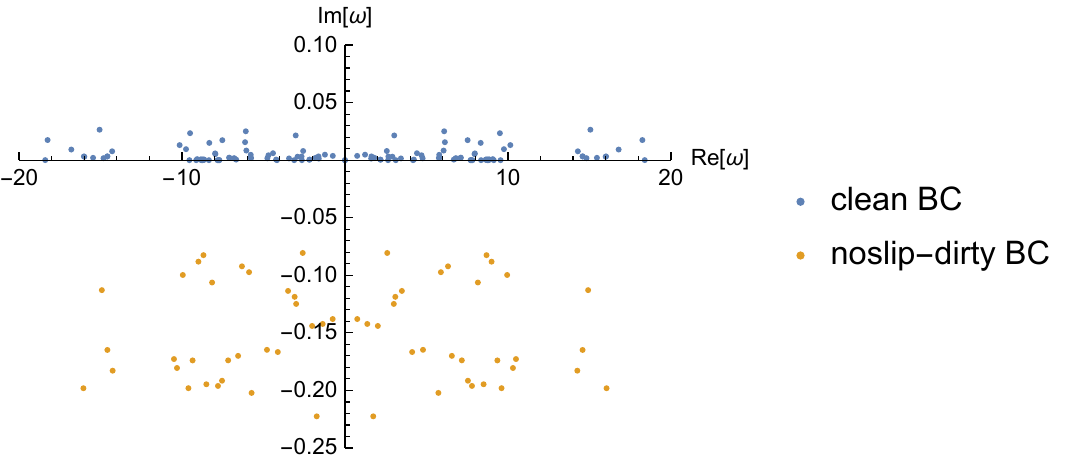}
\caption{The eigenvalue spectrum in the ballistic (collisionless) limit $\tau_{\mathrm{ee}} \to \infty$, at $u_0 = 1/10$ and using the $\theta$ discretization. Blue dots show the spectrum for clean boundary conditions at both ends, and yellow dots for dirty boundary conditions at $x = L$.}
\label{fig:ballistic}
\end{figure}


\section{Experimental Outlook}

In this letter, we have numerically computed $\gamma$ across the ballistic-to-hydrodynamic crossover, in a cavity with the Dyakonov-Shur boundary conditions. We observed that the fate of the instability in the ballistic limit is sensitive to boundary conditions on non-hydrodynamic modes. This provides a further mechanism for suppressing the instability in experimental systems.

The calculations of this paper appear most important for the Fermi liquid of graphene, where $v_{\mathrm{F}} \tau_{\mathrm{ee}} \sim L$ \cite{bandurin, levitov1703}. However, it is believed that other 2DEGs, such as GaAs-based heterostructures, are deeper in the hydrodynamic limit, with $\tau_{\mathrm{imp}} \gtrsim 10\tau_{\mathrm{ee}}$ and $v_{\mathrm{s}}\tau_{\mathrm{ee}} \ll L$ \cite{DS}. However, we observe in Figure \ref{fig:gammafunc} that the hydrodynamic regime (where $\gamma$ is a decreasing function of $\tau_{\mathrm{ee}}$)  shrinks substantially if $v_{\mathrm{g}}\gg v_{\mathrm{F}}$; see also \cite{lucasdassarma}.    If the modes responsible for the DS instability need not be hydrodynamic even if $\tau_{\mathrm{ee}} \ll \tau_{\mathrm{imp}}$, then the hydrodynamic assumption frequently employed in the literature may need scrutiny.  

We suggest a careful study  of electronic boundary conditions in the cavities where the DS instability is searched for, perhaps using transverse electron focusing \cite{dgg}.   This technique has revealed clean boundaries with almost specular reflection in graphene \cite{dgg}.  In a system with clean boundary conditions, our work predicts the DS instability both in a hydrodynamic limit, and in a collisionless limit at very low temperatures where electron-phonon scattering is negligible.    Furthermore, at higher temperatures, the absence of the DS instability could be used as a heuristic \emph{upper and lower bound} on $\tau_{\mathrm{ee}}$ and $\nu$.  Direct probes of $\nu$ are challenging \cite{tomadin}, and indirect measures are imprecise \cite{polini, levitovhydro}. Another measure of $\tau_{\mathrm{ee}}$ and $\nu$ will prove useful for matching theories of electronic hydrodynamics to experiments.

\addcontentsline{toc}{section}{Acknowledgements}
\section*{Acknowledgements}
We thank Sankar Das Sarma, Kin Chung Fong and Marco Polini for useful discussions.
CBM was supported by the U.S.\ Department of Energy, Office of Basic Energy Sciences, Division of Materials Sciences and Engineering, under Contract No.~DE-AC02-76SF00515, and the Alexander von Humboldt Foundation via a Feodor Lynen fellowship. AL was supported by the Gordon and Betty Moore Foundation's EPiQS Initiative through Grant GBMF4302.

\begin{appendix}

\section{Nonlinear Hydrodynamics from Kinetic Theory}
\label{app:kinetic}

This appendix derives explicitly the kinetic theory used in the main text.

\subsection{Collision Integral}
We begin by assuming a 2d Fermi liquid with a single band with dispersion relation \eqref{eq:gal}. The full kinetic equation is \begin{equation}
\partial_t f + \frac{p_i}{m} \frac{\partial f}{\partial x_i} + F_i^{\mathrm{ext}} \frac{\partial f}{\partial p_i} = \mathcal{C}[f]
\label{eq:boltzmann_abstract}
\end{equation}
where $\mathcal{C}[f]$ is the nonlinear collision integral, and $F_i^{\mathrm{ext}}$ are external forces which we will, for now, set to zero. We now postulate an ansatz for a nonlinear function $\mathcal{C}$ which is manifestly consistent with local charge and momentum conservation. Following the relaxation time approximation \cite{bgk}, we write \begin{equation}
 \mathcal{C}[f] = -\frac{f-F_{\mathrm{eq}}[f]}{\tau_{\mathrm{ee}}},
\label{eq:bgk}
\end{equation}
where \begin{equation}
F_{\mathrm{eq}}[f] \equiv \mathrm{\Theta}\left(\mu[f] + {\mathbf{u}}[f]\cdot \mathbf{p} - \epsilon(\mathbf{p})\right),
\end{equation}
with $\tilde \mu$ and $\tilde{\mathbf{u}}$ determined self-consistently by the equations
\begin{subequations}\label{eq:consC}
\begin{align}
\int \frac{\mathrm{d}^d\mathbf{p}}{(2\mpi \hbar)^d} \; f &= \int \frac{\mathrm{d}^d\mathbf{p}}{(2\mpi \hbar)^d} \; F_{\mathrm{eq}}( \mu[f], \mathbf{u}[f]) \equiv n, \\
\int \frac{\mathrm{d}^d\mathbf{p}}{(2\mpi \hbar)^d} \; f\mathbf{p} &= \int \frac{\mathrm{d}^d\mathbf{p}}{(2\mpi \hbar)^d} \; F_{\mathrm{eq}}( \mu[f], {\mathbf{u}}[f])\mathbf{p}.
\end{align}
\end{subequations}
Combining \eqref{eq:boltzmann_abstract}, \eqref{eq:bgk} and \eqref{eq:consC}, it is straightforward to see that globally, charge and momentum are conserved. We have also defined the local density $n(\mathbf{x},t)$ above. The choice of the equilibrium distribution $F_{\mathrm{eq}}[f]$ is motivated as an approximation of the zero temperature limit of the Fermi-Dirac distribution. Due to the simple dispersion relation \eqref{eq:gal}, it is possible to explicitly evaluate \eqref{eq:consC}. We find, suitably shifting the integration variable via $\mathbf{q} = \mathbf{p} - m \mathbf{u}$:
\begin{subequations}
\begin{align}
n = \int \frac{\mathrm{d}^2\mathbf{p}}{(2\mpi \hbar)^2} \; \mathrm{\Theta}\!\left(\mu + \mathbf{p} \cdot \mathbf{u} - \frac{\mathbf{p}^2}{2m}\right) &= \int \frac{\mathrm{d}^2\mathbf{q}}{(2\mpi \hbar)^2} \; \mathrm{\Theta}\!\left(\mu + \frac{m\mathbf{u}^2}{2} - \frac{\mathbf{q}^2}{2m}\right) = \frac{(m\mathbf{u})^2 + 2m\mu}{4\mpi \hbar^2}, \\
\int \frac{\mathrm{d}^2\mathbf{p}}{(2\mpi \hbar)^2} f\mathbf{p} &= \int \frac{\mathrm{d}^2\mathbf{p}}{(2\mpi \hbar)^2} (\mathbf{q}+m\mathbf{u}) \mathrm{\Theta}\!\left(\mu + \frac{m\mathbf{u}^2}{2} - \frac{\mathbf{q}^2}{2m}\right)= mn\mathbf{u}.
\end{align}
\end{subequations}
Inverting these relations: \begin{subequations}\begin{align}
\mathbf{u}[f] &= \frac{\int\mathrm{d}^d\mathbf{p} \; f\mathbf{p}}{m\int\mathrm{d}^d\mathbf{p} \; f}, \\
\mu[f] &= \frac{1}{2\mpi m}\int \mathrm{d}^2\mathbf{p} f - \frac{1}{2m}\left(\frac{\int\mathrm{d}^d\mathbf{p} \; f\mathbf{p}}{\int\mathrm{d}^d\mathbf{p} \; f}\right)^2 
\end{align}\end{subequations}
Finally, is useful to define an effective Fermi velocity \begin{equation}
v_{\mathrm{F}} = \sqrt{\mathbf{u}^2 + \frac{2\mu}{m}} = \frac{2\mpi \hbar}{m} \sqrt{\frac{n}{\mpi}},
\label{eq:vFn}
\end{equation}
such that all excitations at the Fermi surface obey $|\mathbf{v} - \mathbf{u}| = v_{\mathrm{F}}$. Given $\mathbf{u}[f]$ and $\mu[f]$, we have hence specified the collision integral $\mathcal{C}[f]$.

\subsection{The Navier-Stokes Equations}
It is an instructive exercise to explicitly derive the first-order (in the gradient expansion) hydrodynamic equations from this kinetic theory. Integrating \eqref{eq:boltzmann_abstract} over $\mathbf{p}$, one finds the exact result \begin{equation}
\partial_t n + \partial_i (nu_i) = 0.
\end{equation}
The conservation of charge holds exactly, as it must. Multiplying by $p_i/m$ before integrating, we find an equation for the conservation of momentum:\begin{equation}
\partial_t (nu_i) + \partial_j \int \frac{\mathrm{d}^2\mathbf{p}}{(2\mpi \hbar)^2} \frac{p_i p_j}{m^2} f = 0. \label{eq:momcons}
\end{equation}
The hydrodynamic equations are found by computing the second term above, order-by-order in $\tau_{\mathrm{ee}}$, which is taken to be a small parameter.\footnote{More precisely, we expand in the small parameter $\tau_{\mathrm{ee}}\partial_t$.}

At zeroth order in $\tau_{\mathrm{ee}}$, the collision integral must be identically satisfied: $\mathcal{C}[f]=0$. Hence, $f = F_{\mathrm{eq}}[f]$, and so \begin{equation}
\int \frac{\mathrm{d}^2\mathbf{p}}{(2\mpi \hbar)^2} \frac{p_i p_j}{m^2} F_{\mathrm{eq}} = \int \frac{\mathrm{d}^2\mathbf{q}}{(2\mpi \hbar)^2} \frac{(q_i+mu_i)(q_j+mu_j)}{m^2} \mathrm{\Theta}\!\left(\mu + \frac{m\mathbf{u}^2}{2} - \frac{\mathbf{q}^2}{2m}\right) = nu_i u_j + \frac{P(n)}{m}\mdelta_{ij},
\label{eq:Euler}
\end{equation}
where $P(n)$ is the hydrodynamic pressure per unit mass, given by \begin{equation}
P(n) = (2\mpi \hbar)^2 \frac{n^2 }{4\mpi m}.
\label{eq:Pn}
\end{equation}
Combining \eqref{eq:momcons} and \eqref{eq:Euler} we recover the dissipationless Euler equation.

At first order in $\tau_{\mathrm{ee}}$, we may write $f=f^1 + F_{\mathrm{eq}}$ where \begin{equation}
f^1 = -\tau_{\mathrm{ee}}\left(\partial_t F_{\mathrm{eq}} +\frac{p_i}{m} \partial_i F_{\mathrm{eq}}\right).
\end{equation}
Hence, \begin{equation}
 \int \frac{\mathrm{d}^2\mathbf{p}}{(2\mpi \hbar)^2} \frac{p_i p_j}{m^2} f = n u_i u_j + \frac{P(n)}{m}\mdelta_{ij} + \tau_{ij},
\end{equation}
where $\tau_{ij}$ is a dissipative stress tensor given by \begin{equation}
\tau_{ij} = \int \frac{\mathrm{d}^2\mathbf{p}}{(2\mpi \hbar)^2} \frac{p_i p_j}{m^2}f^1 = -\tau_{\mathrm{ee}}\int \frac{\mathrm{d}^2\mathbf{p}}{(2\mpi \hbar)^2} \frac{1}{m^2} \left[\partial_t (p_i p_j F_{\mathrm{eq}}) + \frac{1}{m} \partial_k (p_i p_j p_k F_{\mathrm{eq}})\right].
\end{equation}
Using the explicit form of $F_{\mathrm{eq}}$ we compute \begin{equation}
\tau_{ij} = -\tau_{\mathrm{ee}}\big[\partial_t \left(nu_i u_j + \frac{P}{m}\mdelta_{ij}\right) + \partial_k \left(nu_i u_j u_k + \frac{P}{m} (u_i \mdelta_{jk} + u_j \mdelta_{ik} + u_k \mdelta_{ij}) \right)\big].
\end{equation}
We may now use the zeroth order equations of motion, along with \eqref{eq:Pn}, to simplify this expression: \begin{equation}
\begin{split}
\tau_{ij} &= -\tau_{\mathrm{ee}} \Big[u_i \partial_t \left(mnu_j\right) + u_j \partial_t \left(mnu_i\right) -m u_i u_j \partial_t n + u_i \partial_k (mnu_j u_k + P(n) \mdelta_{jk}) + u_j \partial_k (mnu_i u_k + P(n) \mdelta_{ik}) \\
& \qquad\quad - u_i u_j \partial_k (mnu_k) + P(\partial_j u_i + \partial_j u_i) + \partial_k(Pu_k)\mdelta_{ij} + \partial_t P \mdelta_{ij} \Big] \\
&= -\tau_{\mathrm{ee}} P \left(\partial_j u_i + \partial_i u_j + \mdelta_{ij} \left(u_k \frac{2\partial_k n}{n} + \partial_k u_k + 2\frac{\partial_t n}{n} \right)\right) \\
&= - \tau_{\mathrm{ee}} P \left(\partial_i u_j + \partial_j u_i - \mdelta_{ij} \partial_k u_k\right).
\end{split}
\end{equation}
This is a viscous stress tensor with vanishing bulk viscosity (consistent with Galilean invariance), and a shear viscosity (per unit mass) \begin{equation}
\eta = \tau_{\mathrm{ee}}P = \frac{m\tau_{\mathrm{ee}}nv_{\mathrm{F}}^2}{4}.
\end{equation}

\subsection{Linearized Kinetic Equations}

Let us now consider a static equilibrium with a non-vanishing velocity $\mathbf{u}_0$. Our goal is to linearize the collision operator. This can be accomplished by brute force:
\begin{equation}
\begin{split}
\mathcal{C}[f] &\approx \frac{1}{\tau_{\mathrm{ee}}}\left[\int \mathrm{d}^2\mathbf{q} \left( \frac{\partial F_{\mathrm{eq}}(\mathbf{p})}{\partial \mu} \frac{\mdelta \mu}{\mdelta f(\mathbf{q})}\mdelta f(\mathbf{q}) + \frac{\partial F_{\mathrm{eq}}(\mathbf{p})}{\partial u_j} \frac{\mdelta u_j}{\mdelta f(\mathbf{q})}\mdelta f(\mathbf{q}) \right) - \mdelta f(\mathbf{p})\right] \\
&= \frac{1}{\tau_{\mathrm{ee}}}\left[\int \mathrm{d}^2\mathbf{q} \, \mdelta\!\left(\mu + \mathbf{p}\cdot \mathbf{u}_0 - \frac{\mathbf{p}^2}{2m}\right) \left( \frac{\mdelta \mu}{\mdelta f(\mathbf{q})}\mdelta f(\mathbf{q}) + p_j \frac{\mdelta u_j}{\mdelta f(\mathbf{q})}\mdelta f(\mathbf{q}) \right) - \mdelta f(\mathbf{p})\right].
\end{split}
\end{equation}
We know that
\begin{subequations}
\begin{align}
 \frac{\mdelta u_i}{\mdelta f(\mathbf{q})} &= \frac{1}{(2\mpi \hbar)^2} \left[ \frac{q_i}{m n} - \frac{u_i}{n}\right], \\
 \frac{\mdelta \mu }{\mdelta f(\mathbf{q})} &= \frac{1}{2\mpi m} -mu_i \frac{\mdelta u_i}{\mdelta f(\mathbf{q})},
\end{align}
\end{subequations}
From the form of the collision operator and the low temperature Fermi-Dirac distribution, it is clear that only the dynamics at the Fermi surface is relevant. Invoking \eqref{eq:Phi}, we obtain \begin{equation}
\mdelta f(\mathbf{p}) = \mdelta \left(\mu + \frac{m}{2}\mathbf{u}^2 - \frac{\tilde{\mathbf{p}}^2}{2m}\right) \sum_{n=-\infty}^\infty a_n(\mathbf{x},t) \mathrm{e}^{\mathrm{i}n\theta}.
\label{eq:deltaf}
\end{equation}
where $\tilde{\mathbf{p}} = \mathbf{p}-m\mathbf{u}$ and $\tan \theta = \tilde p_y/\tilde p_x$. We can now perform the $q$-integrals explicitly in the collision operator. Letting $m v_{\mathrm{F}} \equiv p_{\mathrm{F}}$:
\begin{equation}
\begin{split}
\int \mathrm{d}^2\tilde{\mathbf{q}}\, \mdelta\!\left(\mu + \frac{m}{2}\mathbf{u}^2 - \frac{\tilde{\mathbf{q}}^2}{2m}\right) \mdelta f(\mathbf{q}) \left(\frac{1}{2\mpi m} + \frac{\tilde p_i \tilde q_i}{(2\mpi \hbar^2)mn}\right) &= \int \frac{\mathrm{d}\theta_q}{2\mpi} \left(1 + 2\frac{\tilde p_i \tilde q_i}{p_{\mathrm{F}}^2}\right)\mdelta f(\mathbf{q}) \\
&= a_0 + \frac{\tilde p_x + \mathrm{i}\tilde p_y}{p_{\mathrm{F}}}a_1 + \frac{\tilde p_x - \mathrm{i}\tilde p_y}{p_{\mathrm{F}}}a_{-1},
\end{split}
\end{equation}
we obtain \begin{equation}
\mathcal{C}[f] = - \frac{\mdelta f}{\tau_{\mathrm{ee}}} + \frac{1}{\tau_{\mathrm{ee}}} \mdelta\left(\mu + \frac{m}{2}\mathbf{u}^2 - \frac{\tilde{\mathbf{p}}^2}{2m}\right) \sum_{|n|\le 1} \mathrm{e}^{\mathrm{i}n\theta} a_n,
\end{equation}
which is equivalent to the collision integral in \eqref{eq:boltzmann} and \eqref{eq:PPhi}, even when $\mathbf{u}_0 \ne \mathbf{0}$. Indeed, the full Boltzmann equation now reads \begin{equation}
\partial_t \mdelta f + \mathbf{v}(\mathbf{p})\cdot \frac{\partial \mdelta f}{\partial \mathbf{x}} = \mathcal{C}[f],
\label{eq:A3boltz}
\end{equation}
and integrating over the magnitude $|\tilde p|$, we obtain \begin{equation}
\partial_t \Phi + \big(u_0 + v_{\mathrm{F}}\cos(\theta)\big) \partial_x \Phi = -\frac{1}{\tau_{\mathrm{ee}}} \mathsf{P}[\Phi].
\end{equation}

\subsection{External Forces from a Gate}\label{app:gate}

We now address the role of external forcing, given in \eqref{eq:Fgate}. We stick to the linear response limit. Using \eqref{eq:deltaf} we find \begin{equation}
\mdelta n = \int \frac{\mathrm{d}^2\mathbf{p}}{(2\mpi \hbar)^2} \mdelta f = \frac{m}{2\mpi \hbar^2} a_0,
\label{eq:na0}
\end{equation}
and so we see that the external force is given by \begin{equation}
\mathbf{F} = \frac{e^2m}{2\mpi \hbar^2C} \nabla a_0.
\end{equation}
In the Boltzmann equation, $\mathbf{F}$ is only non-vanishing within linear response, and so when computing $\mathbf{F} \cdot \partial f/\partial \mathbf{p}$ we may approximate $f\approx f_{\mathrm{eq}}$. Hence, we find \begin{equation}
\mathbf{F} \cdot \frac{\partial f}{\partial \mathbf{p}} = -\mdelta \left(\mu + \mathbf{u}_0 \cdot \mathbf{p} - \frac{\mathbf{p}^2}{2m}\right) \left(\mathbf{u}_0 - \mathbf{v}(\mathbf{p})\right) \cdot \frac{e^2m}{2\mpi \hbar^2C} \nabla a_0 = \mdelta \left(\mu + \mathbf{u}_0 \cdot \mathbf{p} - \frac{\mathbf{p}^2}{2m}\right) \frac{e^2mv_{\mathrm{F}}}{2\mpi \hbar^2C} \cos(\theta) \partial_x a_0.
\end{equation}
In the last step we have made the simplifying assumption for this paper that all spatial dependence is constrained to the $x$ direction, and used the fact that $\mathbf{v} - \mathbf{u}_0$ at the Fermi surface is constrained to vectors of length $v_{\mathrm{F}}$. Now, using the relations in \eqref{eq:vgate} and \eqref{eq:vFn}, we find \begin{equation}
\mathbf{F} \cdot \frac{\partial f}{\partial \mathbf{p}} = \mdelta \left(\mu + \mathbf{u}_0 \cdot \mathbf{p} - \frac{\mathbf{p}^2}{2m}\right) \frac{2n_0e^2}{mCv_{\mathrm{F}}} \cos(\theta) \partial_x a_0 = \mdelta \left(\mu + \mathbf{u}_0 \cdot \mathbf{p} - \frac{\mathbf{p}^2}{2m}\right) \frac{2v_{\mathrm{g}}^2}{v_{\mathrm{F}}} \cos(\theta) \partial_x a_0
\end{equation}
As this term is added to the left hand side of \eqref{eq:A3boltz}, after integrating over $|\tilde p|$, we obtain \eqref{eq:boltzmannfin}.

\subsection{Hydrodynamic Modes and the Viscosity} \label{app:hydromode}
In the hydrodynamic limit, the only relevant components of the kinetic equations (with gate effects accounted for) are $a_n$ for $|n|\le 2$ \cite{lucas1612}. The relevant components of \eqref{eq:boltzmannfin} read
\begin{subequations}
\label{eq:appmodes}
\begin{align}
\frac{\partial a_0}{\partial t} + u_0 \frac{\partial a_0}{\partial x} + \frac{v_{\mathrm{F}}}{2}\left( \frac{\partial a_1}{\partial x} + \frac{\partial a_{-1}}{\partial x}\right) &= 0, \label{eq:appmodesA} \\
\frac{\partial a_{\pm 1}}{\partial t} + u_0 \frac{\partial a_{\pm 1}}{\partial x} + \frac{v_{\mathrm{F}}}{2}\left( \frac{\partial a_{0}}{\partial x} + \frac{\partial a_{\pm 2}}{\partial x}\right) + \frac{v_{\mathrm{g}}^2}{v_{\mathrm{F}}} \frac{\partial a_0}{\partial x} &= 0, \\
\frac{\partial a_{\pm 2}}{\partial t} + \frac{v_{\mathrm{F}}}{2} \partial_x a_{\pm1} &= -\frac{a_{\pm 2}}{\tau_{\mathrm{ee}}}. \label{eq:appmodelast}
\end{align}
\end{subequations}
We will then look for plane wave solutions proportional to $\mathrm{e}^{\mathrm{i}(k x-\omega t)}$.

In the hydrodynamic limit $\omega \tau_{\mathrm{ee}} \ll 1$, we can approximate that $a_{\pm 2} \approx -\frac{1}{2}\tau_{\mathrm{ee}}v_{\mathrm{F}} \partial_x a_{\pm 1}$. We define \begin{equation}
a_\pm = \frac{a_1 \pm a_{-1}}{2}.
\end{equation}
One then finds the equation \begin{equation}
-\mathrm{i}(\omega - u_0k) a_- + \frac{v_{\mathrm{F}}^2\tau_{\mathrm{ee}}}{4} k^2 a_- = 0.
\end{equation}
This is the shear diffusive mode for momentum, in a background velocity field: \begin{equation}
\omega = u_0 k - \mathrm{i} \nu k^2.
\end{equation}
Recall the definition of $\nu$ in \eqref{eq:nudef}. Of more interest to us are the sound modes, which couple together $a_0$ and $a_+$: \begin{subequations}\begin{align}
-(\omega - u_0k)a_0 + v_{\mathrm{F}} k a_+ &= 0, \\
-\mathrm{i}(\omega - u_0k)a_+ + \frac{v_{\mathrm{s}}^2}{2v_{\mathrm{F}}} \mathrm{i}ka_0 + \frac{v_{\mathrm{F}}^2\tau_{\mathrm{ee}}}{4} k^2 a_+ &= 0
\end{align}\end{subequations}
with $v_{\mathrm{s}}$ defined in \eqref{eq:nudef}. These equations can be solved in the limit $\omega \rightarrow 0$, and they yield approximately \begin{equation}
\omega = (u_0 \pm v_{\mathrm{s}})k - \mathrm{i}\frac{v_{\mathrm{F}}^2\tau_{\mathrm{ee}}}{8} k^2 + \cdots
\end{equation}
At $\mathrm{O}(k^3)$ corrections to this equation will arise from the terms that we neglected in \eqref{eq:appmodelast}.

A very crude truncation of the kinetic equations on time scales comparable to (or shorter than) $\tau_{\mathrm{ee}}$ is to replace \eqref{eq:boltzmannfin} with \eqref{eq:appmodes}, and not to treat any terms as small or large relative to others. In this approximation, we find that the only change to the hydrodynamic modes above is that $\nu$ is replaced by $\nu(\omega)$, as given in \eqref{eq:nuomega}.



\section{Numerical Methods}
\label{app:numerics}

To numerically obtain the spectrum, we make use of the quasi one-dimensional structure (with respect to $x \in [0, L]$) of the problem. 

\subsection{Harmonic Moment Representation}

The Boltzmann equation \eqref{eq:boltzmannfin} represented by the harmonic coefficients $a_n(x,t) = \mathrm{e}^{-\mathrm{i}\omega t} a_n(x)$ in \eqref{eq:Phi} reads
\begin{equation}
\label{eq:boltzmann_harmonic}
-\mathrm{i}\omega\,a_n(x) + u_0\,a_n'(x) + \frac{v_{\mathrm{F}}}{2} \left( a_{n-1}'(x) + a_{n+1}'(x) \right) + \frac{v_{\mathrm{g}}^2}{v_{\mathrm{F}}} \mdelta_{|n|,1}\,a_0'(x) = W_n a_n(x),
\end{equation}
where $x \in [0, L]$, and
\begin{equation}
W_n = \begin{cases} -1/\tau_{\mathrm{ee}} & |n| \ge 2 \\ 0 & \text{otherwise} \end{cases}.
\end{equation}
In the numerical implementation, we choose a cut-off $n_{\max}$ and truncate $a_n(x) = 0$ for $|n| > n_{\max}$. Unless specified otherwise, $n_{\max} = 12$. We have checked that this value is sufficient for an accurate determination of the spectrum. Then Eq.~\eqref{eq:boltzmann_harmonic} is formally solved by
\begin{equation}
\vec{a}(x) = \mathrm{e}^{A x} \, \vec{a}(0)
\end{equation}
with $\vec{a}(x) = (a_{-n_{\max}}(x), a_{-n_{\max}+1}(x), \dots, a_{n_{\max}}(x))^{\mathsf{T}}$ the vector of $a_n$'s and the matrix (assuming $u_0 \neq 0$)
\begin{equation}
\label{eq:Adef}
A = \begin{pmatrix}
\ddots & \ddots & & & & & \\
\ddots & u_0 & \frac{v_{\mathrm{F}}}{2} & & & & \\
& \frac{v_{\mathrm{F}}}{2} & u_0 & \frac{v_{\mathrm{F}}}{2} + \frac{v_\mathrm{g}^2}{v_\mathrm{F}} & & & \phantom{\ddots} \\
& & \frac{v_{\mathrm{F}}}{2} & u_0 & \frac{v_{\mathrm{F}}}{2} & & \phantom{\ddots} \\
& & & \frac{v_{\mathrm{F}}}{2} + \frac{v_\mathrm{g}^2}{v_\mathrm{F}} & u_0 & \frac{v_{\mathrm{F}}}{2} & \phantom{\ddots} \\
& & & & \frac{v_{\mathrm{F}}}{2} & u_0 & \ddots \\
& & & & & \ddots & \ddots \\
\end{pmatrix}^{-1}
\begin{pmatrix}
\ddots & & & & & & \\
& \mathrm{i} \omega - \frac{1}{\tau_{\mathrm{ee}}} & & & & & \\
& & \mathrm{i} \omega \quad & & & & \\
& & & \mathrm{i} \omega \quad & & & \\
& & & & \mathrm{i} \omega \quad & & \\
& & & & & \mathrm{i} \omega - \frac{1}{\tau_{\mathrm{ee}}} & \\
& & & & & & \ddots
\end{pmatrix}.
\end{equation}
In practice, we diagonalize $A$ and determine its eigenvalues $\lambda_j$ and corresponding eigenvectors $\vec{\psi}_j$, i.e.,
\begin{equation}
A \vec{\psi}_j = \lambda \vec{\psi}_j \quad\text{for}\quad j = 1, 2, \dots 2n_{\max}+1.
\end{equation}
Thus
\begin{equation}
\vec{a}(x) = \sum_{j=1}^{2n_{\max}+1} c_j \, \mathrm{e}^{\lambda_j (x - \delta_j L)} \vec{\psi}_j
\end{equation}
with coefficients $c_j$ to be determined by the boundary conditions, and
\begin{equation}
\delta_j = \begin{cases} 1 & \mathrm{Re}(\lambda_j) > 0 \\ 0 & \mathrm{Re}(\lambda_j) \le 0 \end{cases}.
\end{equation}
The term $\delta_j L$ ensures the numerically advantageous property $|\mathrm{e}^{\lambda_j (x - \delta_j L)}| \le 1$ for all $x \in [0, L]$.

For some boundary conditions, including the clean boundary conditions we emphasized in the main text, a further simplification is possible. By adding the equations \eqref{eq:boltzmann_harmonic} for the modes $a_n$ and $a_{-n}$ together, we find a closed, reduced set of differential equations for $a_0$, $a_1 + a_{-1}$, $a_2 + a_{-2}, \ldots, a_{n_{\mathrm{max}}} + a_{-n_{\mathrm{max}}} $. This allows us to reduce the size of the numerical problem and increase the value of $n_{\max}$.

We represent the left and right boundary conditions via matrices $B_{\mathrm{left}}$ and $B_{\mathrm{right}}$ as $B_{\mathrm{left}}\,\vec{a}(0) = 0$ and $B_{\mathrm{right}}\,\vec{a}(L) = 0$, respectively. Arranging the vectors $\mathrm{e}^{\lambda_j (-\delta_j L)} \vec{\psi}_j$ as columns into a matrix $V_{\mathrm{left}}$ and the vectors $\mathrm{e}^{\lambda_j (L - \delta_j L)} \vec{\psi}_j$ as columns into a matrix $V_{\mathrm{right}}$, the boundary conditions can be written as
\begin{equation}
\label{eq:Bcondition}
B\,\vec{c} = 0, \quad
B = \begin{pmatrix}
B_{\mathrm{left}} V_{\mathrm{left}} \\
\hline 
B_{\mathrm{right}} V_{\mathrm{right}}
\end{pmatrix}
\end{equation}
with $\vec{c}$ the vector of $c_j$ coefficients. Since the total number of boundary conditions should be equal to the number of coefficients in $\vec{a}(x)$, namely $2 n_{\max} + 1$, $B$ is a square matrix. In other words, the condition \eqref{eq:Bcondition} means that $B$ is singular. To compute a point $\omega$ of the spectrum, we numerically search for a (local) root of the smallest (in magnitude) eigenvalue of $B$ using gradient descent.

\subsection{Angular Discretization in the Ballistic Limit}
\label{app:numerics_angular_discretization}

Complementary to the harmonic representation, one may discretize the $\theta$ variable in the Boltzmann equation \eqref{eq:boltzmannfin}, i.e., compute the distribution function at points $\theta_j = 2 \pi j/n_{\max}$ for $j = 0, 1, \dots, n_{\max}-1$. Due to linearity, one expects that the spectrum depends continuously on the collision term $\frac{1}{\tau_{\mathrm{ee}}}\mathsf{P}[\Phi]$, and thus we simply drop it in the ballistic limit $\tau_{\mathrm{ee}} \to \infty$. Then, for the special case $v_g = 0$ and using the notation $\Phi_j(x) = \Phi(x, \theta_j)$, one arrives at
\begin{equation}
\mathrm{i}\omega\,\Phi_j(x) = \big(u_0 + v_{\mathrm{F}}\cos(\theta_j)\big) \partial_x \Phi_j(x).
\end{equation}
This equation is solved by
\begin{equation}
\Phi_j(x) = \Phi_j(0)\, \mathrm{e}^{\frac{\mathrm{i}\omega}{u_0 + v_{\mathrm{F}}\cos(\theta_j)} x},
\end{equation}
assuming $u_0 + v_{\mathrm{F}}\cos(\theta_j) \neq 0$ for all $j$. In particular, the components $\Phi_j(x)$ are only coupled via the boundary conditions.

\section{Boundary Conditions}
\label{app:bc}

In this appendix, we derive the various boundary conditions employed in the main text.

\subsection{How Many Boundary Conditions are There?}\label{app:count}
Following \cite{lucas1612}, we justify the claim that (in the harmonic basis) the number of boundary conditions is $2n_{\mathrm{max}}$ when $u_0=0$, and $2n_{\mathrm{max}}+1$ when $u_0\ne 0$. Consider\footnote{If $n_{\mathrm{max}}$ is odd, then there will also be two terms in this sum coming from the $v_{\mathrm{g}}$-dependent terms in \eqref{eq:boltzmann_harmonic}. Both of these terms cancel as well.}
\begin{equation}
\begin{split}
\sum_{j=-n_{\mathrm{max}},2-n_{\mathrm{max}},\ldots,n_{\mathrm{max}}} &(-1)^{(j+n_{\mathrm{max}})/2} W_n a_n = (-\mathrm{i}\omega + u_0\partial_x) \left(a_{-n_{\mathrm{max}}} - a_{2-n_{\mathrm{max}}} + \cdots \pm a_{n_{\mathrm{max}}}\right) \\
&+ \frac{v_{\mathrm{F}}}{2} \left(a_{1-n_{\mathrm{max}}}^\prime - (a_{1-n_{\mathrm{max}}}^\prime + a_{3-n_{\mathrm{max}}}^\prime) + \cdots \mp (a_{n_{\mathrm{max}}-3}^\prime + a_{n_{\mathrm{max}}-1}^\prime) \pm a_{n_{\mathrm{max}}-1}^\prime\right).
\end{split}
\end{equation}
Notice that the telescoping sum on the second line above vanishes. If $u_0=0$, we are then left with a \emph{constraint} equation relating the $a_n$s to each other, and so not all $a_n$s are independent. If $u_0\ne 0$, then this equation becomes a differential equation and is no longer a constraint.

What this implies is that the $u_0 \rightarrow 0$ limit is somewhat subtle, as noted in the main text. We must choose boundary conditions which do not become pathological in the $u_0 \rightarrow 0$ limit. In the hydrodynamic limit $\tau_{\mathrm{ee}} \rightarrow 0$, \eqref{eq:boltzmann_harmonic} approximately reduce to \eqref{eq:appmodes}. Since \eqref{eq:boltzmann_harmonic} are first order equations, we should only impose boundary conditions on $a_n$ directly. Noting that when $u_0=0$, \eqref{eq:appmodesA} and \eqref{eq:DSBC} fix \begin{equation}
a_1^\prime(0) + a_{-1}^\prime(0)=0,
\end{equation}
we find from \eqref{eq:appmodelast} that \begin{equation}
a_2(0) + a_{-2}(0)=0.
\label{eq:a2BC}
\end{equation}
Imposing the boundary condition \eqref{eq:a2BC} is redundant when $u_0=0$, and does not lead to an overdetermined problem. When $u_0 \ne 0$, this boundary condition will suppress the hydrodynamic modes that are singular in the $u_0 \rightarrow 0$ limit. This is the boundary condition that we have employed in our simulations, as noted in the main text.

\subsection{Boundary Conditions for the Remaining Degrees of Freedom}

\subsubsection{Clean Boundaries in the Harmonic Representation}

Let us begin by assuming that the boundary conditions are clean. Suppose that everywhere in the box, we had the boundary conditions that $\Phi(\theta) = \Phi(\mpi - \theta)$. Using \eqref{eq:Phi}, we see that \begin{equation}
\Phi(\mpi - \theta) = \sum_{n\in\mathbb{Z}} a_n (-1)^n \mathrm{e}^{-\mathrm{i}n\theta} = \sum_{n\in\mathbb{Z}} (-1)^n a_{-n} \mathrm{e}^{\mathrm{i}n\theta}.
\label{eq:Phimpi}
\end{equation}
Thus we conclude that clean boundary conditions will try to impose \begin{equation}
a_n = (-1)^n a_{-n}.
\label{eq:cleanBC}
\end{equation}
Using the even-odd decomposition of the $a_n$s described above, we conclude that $a_n$ is unconstrained if $n$ is even, and $a_n=0$ if $n$ is odd. When $n_{\mathrm{max}}$ is even, we obtain $n_{\mathrm{max}}$ boundary conditions per boundary.

Obviously, we must slightly modify these boundary conditions for consistency with \eqref{eq:DSBC}. This is relatively simple, because \eqref{eq:cleanBC} decouples different $|n|$. It is sufficient to modify only the $|n|\le 2$ sector. We take \begin{equation}
0 = a_2(0) - a_{-2}(0) = a_2(L) - a_{-2}(L) = a_2(0) + a_{-2}(0),
\label{eq:cleanBC2}
\end{equation}
and also no longer require $a_1(0,L)$ to vanish. \eqref{eq:DSBC} gives us two more boundary conditions, and together with \eqref{eq:cleanBC2}, and \eqref{eq:cleanBC} for $|n|>2$, we find a complete set of $2n_{\mathrm{max}}+1$ boundary conditions. The resulting eigenvalue problem is well-posed for all $\tau_{\mathrm{ee}}$ and $u_0$. 

If $n_{\mathrm{max}}$ is odd, then we have too many boundary conditions. This is why we have only used truncations with $n_{\mathrm{max}}$ even.

\subsubsection{Dirty Boundaries for the Angular Discretization}
\label{app:dirty_bc_theta}

As mentioned above, we use the angular discretization in the ballistic limit, and additionally probe ``dirty'' boundary conditions at $x = L$ (see Fig.~\ref{fig:ballistic} for the corresponding spectrum). For these boundary conditions, the outgoing distribution function (from the boundary) should be featureless, i.e., as uniform as possible, while still satisfying the Dyakonov-Shur current conservation. We partition the angles $\theta_j$ according to positive (ingoing) and negative (outgoing) velocities, i.e., $v_j \ge 0$ or $v_j < 0$ with $v_j \equiv u_0 + v_{\mathrm{F}}\cos(\theta_j)$. Denoting the corresponding index sets by $J_{\mathrm{pos}}$ and $J_{\mathrm{neg}}$, the dirty boundary condition at $x = L$ can be written as
\begin{equation}
\label{eq:dirtyBC_Phi}
\Phi_j(L) = - \frac{1}{\sum_{j' \in J_{\mathrm{neg}}} v_{j'}} \sum_{j'' \in J_{\mathrm{pos}}} v_{j''}\, \Phi_{j''}(L)
\end{equation}
for all $j \in J_{\mathrm{neg}}$, i.e., all ``outgoing'' $\Phi_j$'s have the same value. The condition \eqref{eq:dirtyBC_Phi} directly implies the current conservation
\begin{equation}
\sum_{j} v_j\, \Phi_j(L) = 0,
\end{equation}
where the sum now runs over all discretized angles.

\end{appendix}

\bibliographystyle{unsrt}
\addcontentsline{toc}{section}{References}
\bibliography{dsinstbib}

\end{document}